\documentclass[12pt]{article}
\usepackage{a4,graphicx}
\pdfoutput=1
\begin{document}

\title{Energy resolution improvement in room-temperature CZT detectors}

\author{Y. Ramachers\thanks{E-mail:
  y.a.ramachers@warwick.ac.uk} and D.Y. Stewart\thanks{Email:
  d.y.stewart@warwick.ac.uk}\\Dept. of Physics, University of
  Warwick, Coventry CV4 7AL, UK}
\date{}
\maketitle
%\abstract{We present methods to improve the energy resolution of single
\begin{abstract}We present methods to improve the energy resolution of single
  channel, 
room-temperature Cadmium-Zinc-Telluride (CZT) detectors. A new preamplifier
design enables the
acquisition of the actual transient current from the crystals and
straightforward data 
analysis methods yield unprecedented energy resolution for our
test-detectors. These 
consist of an eV-CAPture Plus crystal as standard and 1 cm cube
Frisch collar
crystals created in-house from low-grade coplanar grid detectors. Energy
resolutions of 1.9\% for our collar detectors and 0.8\% for
the eV crystal at 662 keV were obtained. The latter compares favourably to the
best existing energy resolution results from pixel detectors. 
\end{abstract}
% \keywords{Gamma detectors, Front-end electronics for detector readout, Data
%   processing methods}
\section{Introduction}
The motivation to research CdZnTe (CZT) detectors originates from
participation in the 
COBRA experiment \cite{cobra}, a proposed massive (several hundred kg) array of
CZT crystals for double-beta decay research. Taking into account a typical
mass of merely a few grams for each crystal, several tens of thousands of
crystals will eventually have to be operated reliably over several
years. Naturally, such a set-up would become vastly more practical by
utilising simple
ways to mount and operate individual crystals in the array. 
\par
Single-channel readout for each crystal as opposed to coplanar grid
detectors is considered to be an attractive option. Already the reduction of
wiring close to the crystals by a factor 
of two would be highly 
significant in this case. Frisch collar detectors \cite{collar} appeared to be
the most practical way 
forward, so we modified three existing coplanar grid crystals, low-grade from
eV-Products\footnote{Purchased for the COBRA experiment and kindly provided by
 our collaboration partners.}. In order to have a standard to compare with, we
purchased one 
eV-CAPture Plus technology detector which is of much higher quality \cite{capture}.
\section{Experiment}
\begin{figure}
\begin{center}
\includegraphics[width=12cm]{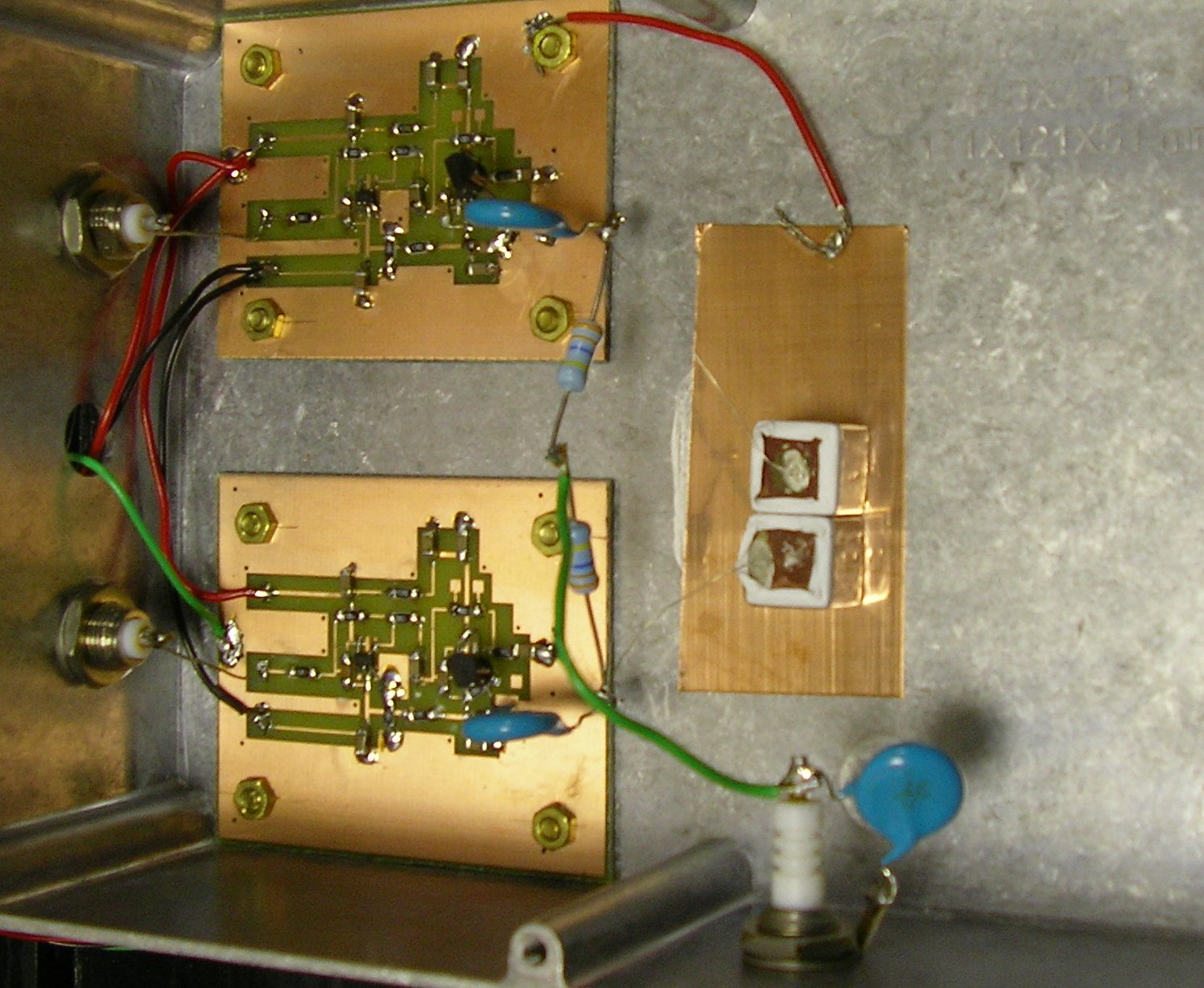}
\caption{Picture of the set up, containing two preamplifiers either for
  readout of two of our in-house Frisch collar crystals (shown here) or
  two-channel readout of anode and cathode signals
  simultaneously.\label{photo}} 
\end{center}
\end{figure}

\begin{figure}
\begin{center}
\includegraphics[width=13cm]{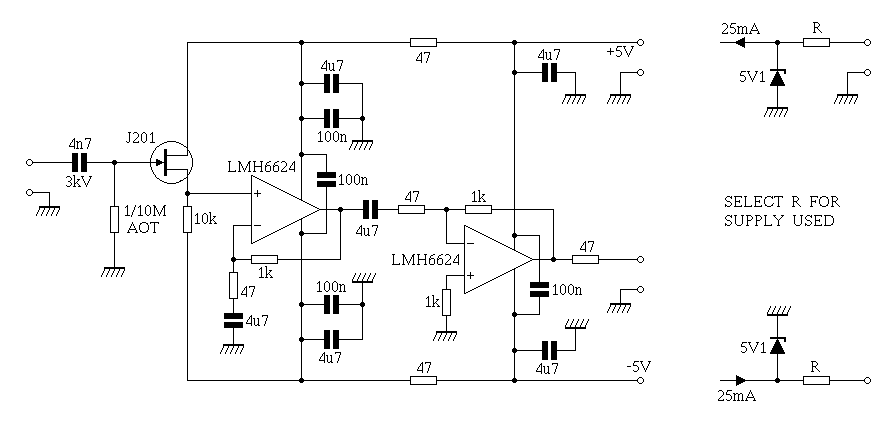}
\caption{The pre-amplifier circuit diagram used in this work. The JFET J201
  can in principle be replaced with any FET input-stage, however low
  intrinsic noise is the crucial selection criterium. The 3kV capacitor at the
input represents the anode AC-coupling.\label{preamp}}
\end{center}
\end{figure}

Crystals and preamplifier are housed together in a standard diecast enclosure
featuring 
connectors for preamplifier power, BNC output and High-voltage input (see
figure \ref{photo}). A single
HV-power supply (Ortec 659, NIM module) delivers both polarities up to
5kV. Two linear DC power supplies deliver $\pm{}5$V to the preamplifier (see
figure \ref{preamp}). The output signal is connected
directly to the data
acquisition system (DAQ) using a 50$\Omega$ BNC cable (RG58). The DAQ consists
of a 100 MHz sampling digital oscilloscope in a 3U PXI module from National
Instruments (NI PXI-5112) mounted in a PXI crate (NI PXI-1042) and is
controlled by an embedded controller PC (NI PXI-8186) running custom-made
LabView software for digital pulse acquisition. Pulses are streamed directly
to disk in binary format for maximum sampling speed when using radioactive
sources for detector calibration. 
\par
The dynamic range is limited to 7-bit (8-bit
oscilloscope, both polarities measured), hence each acquisition needs a little
fine-tuning for the vertical range to capture all important structures. We
emphasize this point since this has become an issue when calibrating with a
Ba-133 source, see figure \ref{ba133}. The 81 keV line resolution is
limited by digitisation noise rather than our analysis method whilst the line
voltage amplitude is simply too small when the system is set-up to acquire the
356 keV line simultaneously. 
\par 
The core of the data acquisition system is the preamplifier design, see
figure \ref{preamp}. It represents a rather unusual method for readout of
semiconductor pulses since it is not a charge-sensitive amplification
system. This preamplifier can be described as a
straightforward source-follower. The motivation to try this type initially 
was to learn more about signal formation in the semiconductor,
i.e. to specialise the analysis and measurement on pulse-shape in contrast to
pulse-height.
\par
The preamplifier is optimised for high-speed, consistent with the data
acquisition bandwidth, and highest possible signal--to--noise ratio. Note that
both properties are almost mutually exclusive, i.e. speeding up the amplifier
worsens the signal and better signal--to--noise ratio slows the
amplifier. However, for applications that require a higher signal gain,
focussing on low-energy signals for example, it is possible to increase the
gain without losing an equal factor in speed. Our application for this
set-up, however, targets rather 
higher energy signals of up to several MeV, hence the circuit in
figure \ref{preamp} is expected to work optimally for us.
\par
The existing three coplanar grid detectors were modified to work as
Frisch collar detectors. All three are cubes of volume 1cm$^{3}$ with gold-plated
electrodes, a full area cathode and a coplanar grid anode structure. All
faces (excluding the cathode face) are covered with insulating paint. Since we were
not interested in operating the grid, paint covering the anode was partly
removed (by dissolving it in ethanol) and the area used to contact
the anode with a wire was attached by a generous drop of silver conductive
paint. The preamplifier is AC-coupled to the crystal anode which is biased
positively by the HV-power supply. The cathode is kept at ground
potential. This
mode was used to operate and test the crystals as simple planar detectors. 
\par
For
the Frisch collar mode, each crystal was wrapped in two layers
of thin teflon tape, covering the full height, leaving out anode and cathode,
similar to devices fabricated in \cite{collar}. The teflon layer was wrapped in a metal foil (aluminium kitchen foil worked best for us) and the
foil attached via a small 'lip' to the cathode (using silver
conductive paint). The metal foil
height determines the performance of the Frisch collar detector
\cite{collar}. We achieved best operating performance with 9mm - 9.5mm foil
height. Any higher shield results in stability problems when biasing the
anode since the grounded shield appears to be too close, particularly at the
cube corners. Finally, prepared crystals can be mounted for operation on a
ground plane. We used a small copper-clad piece of printed-circuit board 
connected to the preamplifier ground, see figure \ref{photo}.
\par
\par
The eV-Products CAPture Plus technology crystal is essentially a Frisch collar
detector using a resistively coupled shield instead of a capacitively one as
outlined above. It has, however, one important feature in addition, a
small anode, utilising the 'small pixel' effect \cite{barrett}. 
The 'best of 
both worlds' has been combined with this type of detector in order to achieve
an electron-signal-only operation with a single channel readout. The
surface area of the rectangular crystal is 1cm$^{2}$ and its height is quoted
as 5mm. It arrived factory--certified with an energy resolution of 2.20\% at
the Cs-137 line at 662 keV and 4.40\% at the Co-57 line at 122 keV. According
to \cite{capture}, an energy resolution of 1.5\% at 662 keV has been
achieved under favourable conditions. The
crystal is contacted and biased identically to our modified crystals, except
that this crystal can withstand higher bias than the Frisch collar crystals
(recommended bias is 1.5 kV). Note that an alternative design, utilising the
capacitively coupled shield and a small anode has been realised in
\cite{mixed} with excellent results.
\par
In case anode as well as cathode signals
are required to be measured with two preamplifiers, it is important to raise the
ground plane (PCB) off the diecast box floor to minimise capacitance for the
cathode signals. A 1cm insulating spacer should be sufficient. The
two-channel readout serves to gain further insight into the signal formation
by picking up anode and cathode signal simultaneously for a single event. The
ratio of these signals is expected to yield depth information \cite{He} and
its sum 
could improve energy resolution by boosting the total amount of charge
collected. So far, we utilised this type of readout only for the eV
crystal. Further research into this mode of operation is in progress. 
\par
\par

\section{Results}
Optimising energy resolution for Frisch collar detectors is fully documented
in \cite{collar}. Performance of crystal operation in this mode was
successfully re-produced. However, in this process we noticed that our
absolute energy resolutions were surprisingly good compared to the
literature. Note that our crystals are remarkably large for CZT crystals
operating as Frisch collar detectors and the crystal quality was expected to
be poor compared to spectroscopy-grade crystals. Additionally, our readout was
custom-made for pulse-shape analysis and not specialised for spectroscopy
operation. Obtaining energy information from integrated digitised pulses is
always considered to be inferior to analogue operations (current integration
on a charge-sensitive preamplifier and spectroscopy shaping). This
'common-knowledge' can lead to interesting preamplifier designs combining charge
and current sensitive readout \cite{hamrita}. Taking all of this into
consideration, we were surprised to see our initial energy resolutions
comparable to published results. In order to achieve further improvements the
following data analysis method has been developed.
\par
Frisch collar detector fabrication inevitably leads to high input capacitances
relative to planar crystals. The metal shield wrapped around the bulk of the
crystal functions by collecting induced charges from any current
flowing inside the capacitively coupled crystal. Thereby, it prevents the
majority of 
the signal appearing on the signal electrodes until the charge cloud comes
close to the anode, hence achieving close to electron-only signal
formation. 
\par
\begin{figure}
\begin{center}
\includegraphics[width=10cm]{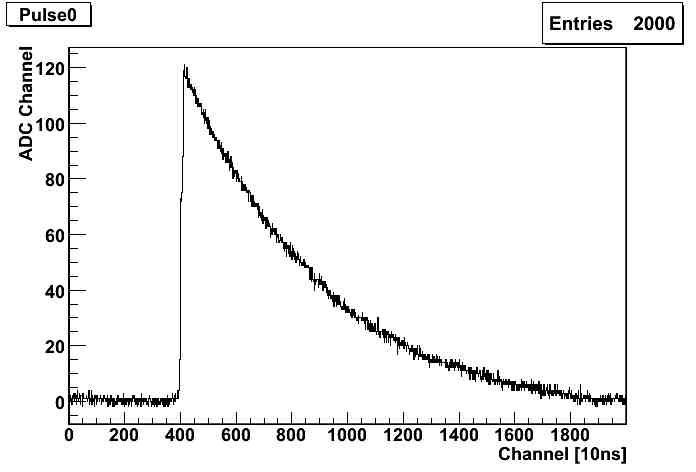}
\caption{Typical output pulse from the preamplifier, digitised with the DAQ
  and streamed to disk directly. The sampling rate was 100 MHz, record length
  2000 channel, vertical range 1.92 Volt. The 'ADC channel' axis corresponds
  to the 
  7-bit digitisation range for positive polarity pulses. This pulse, near the
  maximum range, originates from a 662 keV event.\label{signal1}}
\end{center}
\end{figure}
\par
\begin{figure}
\begin{center}
\includegraphics[width=10cm]{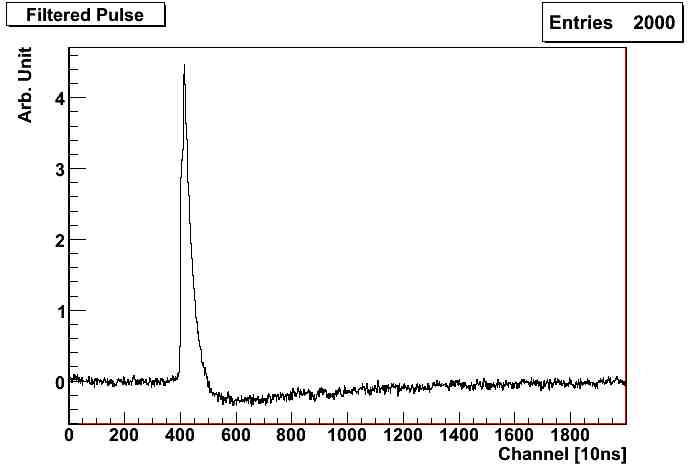}
\caption{The pulse from figure (3) filtered with the software CR-RC filter. \label{signal2}}
\end{center}
\end{figure}
\par
\begin{figure}
\begin{center}
\includegraphics[width=10cm]{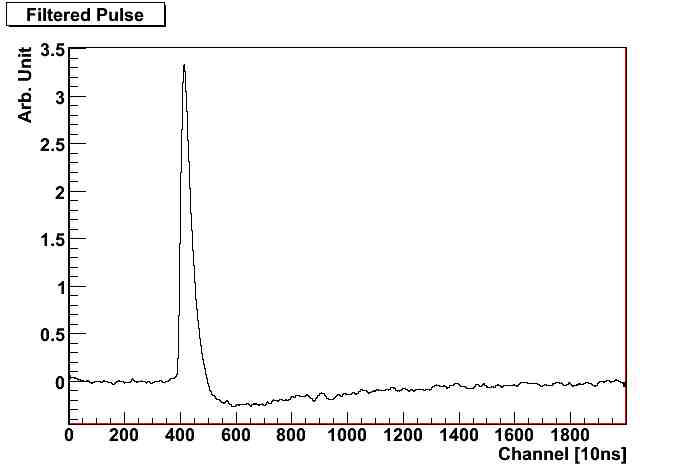}
\caption{The pulse from
  figure (4) but now filtered
  with a moving average filter to remove high-frequency noise but preserving
  the sharp features. This is the final step before analysis of pulse
  parameters commences.\label{signal3}}
\end{center}
\end{figure}
\par
A high input capacitance coupled to the source-follower
preamplifier effectively smears or low-pass--filters the image of the current
flowing in the detector crystal (interpreted here as a current
source). Therefore our signal (see figure \ref{signal1}) is not a true
transient 
current measurement initially. However, a CR-RC filter applied to the
digitised pulse can enhance the fast components in the signal by effectively
transforming the pulse to its rate-of-change image. After shaping, the pulse
more closely represents the actual transient current flowing in the
crystal. We found that for the given preamplifier settings and the data
acquisition bandwidth, time-constants of 200 ns for the high-pass and 4
$\mu$s for the low-pass yield satisfactory results for all our cystals. The
result of such a shaping is shown in figure \ref{signal2}
for a typical pulse. 
\par
Naturally, this shaping process enhances the high-frequency noise, so we
decided to apply a second step to this two-stage software filtering in the
form of a moving
average filter (width 200 ns). Note that a moving average filter belongs to
the class of optimal filters, where the moving average filter
represents the 
optimal method to preserve any steep, fast-changing feature in a pulse while
smoothing high-frequency noise\footnote{All digital filter algorithms as
  described in the text have been
  taken from \protect{\cite{dsp}}.}. Figure \ref{signal3}
shows its effect on
the pulse from figure \ref{signal2}. 
\par
The final analysis step involves calculating basic pulse parameters which
serve to extract the integral and therefore the energy of the pulse. Those
parameters also greatly help in discriminating against corrupt pulses from
pulse pile-up or short-circuit events from over-biasing crystals. We calculate
the following set of parameters: baseline, onset, peak position and amplitude, rise-time,
decay-time and peak broadness (width at 80\% of amplitude left and right
from maximum) and finally the integral and the integration 
risetime. For the two-channel readout we also calculate the sum signal and the
ratio. 
\par
\begin{figure}
\begin{center}
\includegraphics[width=10cm]{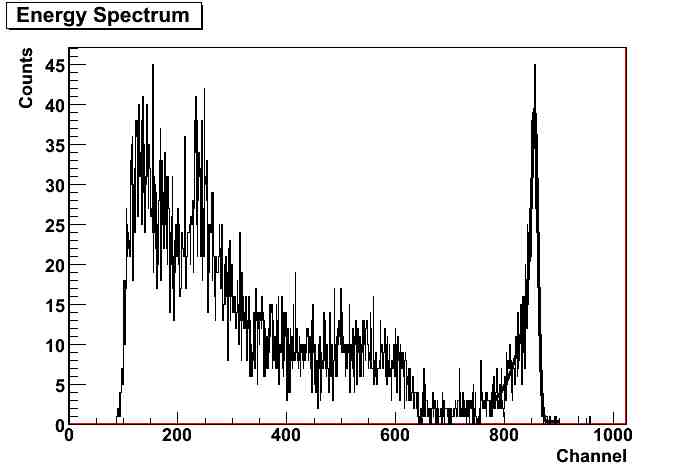}
\caption{Calibration spectrum taken with a Cs-137 source at 1.5kV bias with
  the eV-CAPture Plus detector. Shown is the sum signal from anode and
  cathode, see text.\label{cs137}}
\end{center}
\end{figure}
\par
\begin{figure}
\begin{center}
\includegraphics[width=10cm]{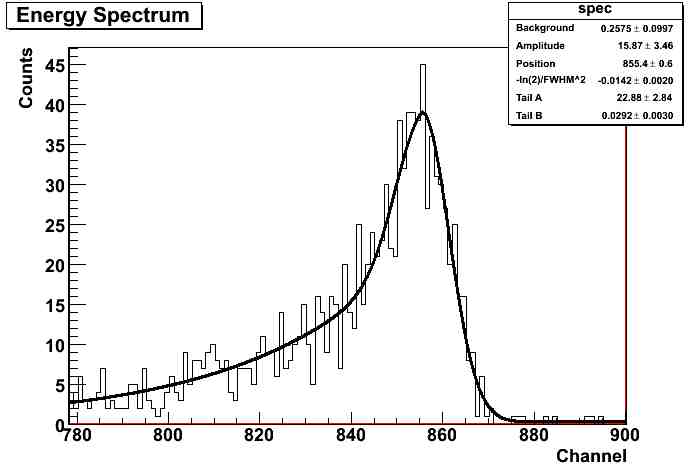}
\caption{Demonstration of the excellent energy resolution achieved with the
  eV-CAPture crystal. Zoom into the spectrum from
  figure (6) around
  the 662 keV line with fitted peak function (solid line on top of histogram).
  Fit parameters with uncertainties are displayed in the legend. The energy
  resolution is 0.82\%. The anode-only signal has 1.04\% (not shown
  here).\label{fit}}
\end{center}
\end{figure}
\par
\begin{figure}
\begin{center}
\includegraphics[width=10cm]{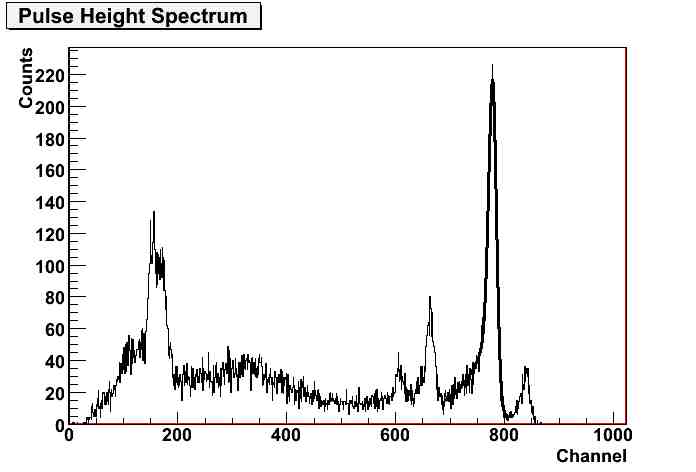}
\caption{Calibration spectrum taken with a collimated Ba-133 source at 1.2kV
  bias with the eV-CAPture Plus technology detector, illuminating the cathode
  side. Note that the 81 keV line energy resolution suffers mainly from
  digitisation noise (limited dynamic range, see discussion in
  section 2).\label{ba133}}
\end{center}
\end{figure}
\par
Figure \ref{cs137} and figure \ref{ba133}
display the energy spectra obtained from
calibrations with a Cs-137 and a Ba-133 source, respectively, both acquired
using our 
test standard, the CAPture Plus crystal. The energy resolutions are listed in
table  \ref{tab1}. They have been obtained from fits to full
energy peaks as
shown in figure \ref{fit} as an example. Due to the inevitable tailing in
single-channel readout systems, we adopted the fitting function from
\cite{arlt}, \cite{ieee} in order to extract reliable energy resolution
values. 
\par
\begin{figure}
\begin{center}
\includegraphics[width=10cm]{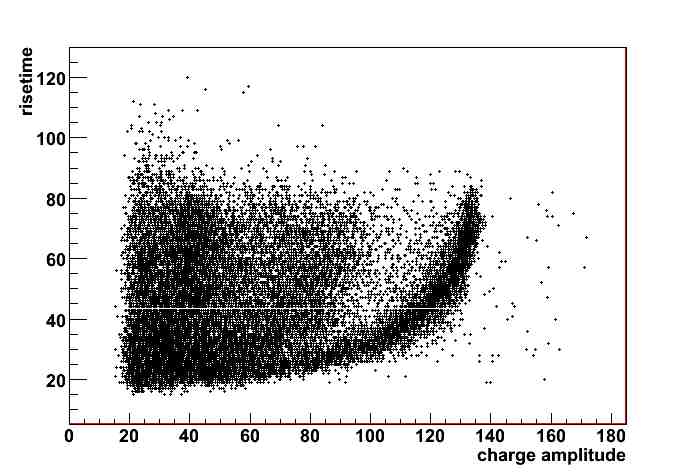}
\caption{Example of a rise-time--pulse height plot (pulse height corresponds
  to charge amplitude, see text) for one of the Frisch collar detectors at 800
  V bias with a 9.5mm collar, irradiated with a Cs-137 source. The energy
  resolution amounts to 1.9\% at 662 keV. As can be seen, a bi-parametric
  correction like demonstrated in \protect{\cite{verger}} would be beneficial
  but could lead to conflicts for our main application, see
  text.\label{rtplot}} 
\end{center}
\end{figure}

At this point it is worth emphasizing that no subsequent data corrections have
been applied, i.e. no rise time to pulse height correlations or any other
correlations have been utilised to improve energy resolution. As shown in
figure \ref{rtplot}, our Frisch collar crystals would benefit
greatly from such an offline correction, whereas the CAPture Plus detector does
not suffer from a similarly strong tailing, mainly due to the small pixel
effect. Ref.~\cite{verger} shows an impressive effect of such a bi-parametric
correction, achieving a similar resolution (less than 1\% at 662 keV) using a
custom-made Frisch collar crystal, to what we achieve with the CAPture Plus
crystal. However, we 
point out that our main application of CZT detectors in fundamental
rare event research (see introduction) does not allow for unspecified
event efficiencies. Any data cut would have to be known with highest possible
precision, hence it is generally better in such an application not to apply
any cut (or energy-dependent corrections) at all. Our analysis merely removes
clearly invalid events such as empty baselines and pulses that saturate the data
acquisition system. 
\par
\par
\begin{figure}
\begin{center}
\includegraphics[width=10cm]{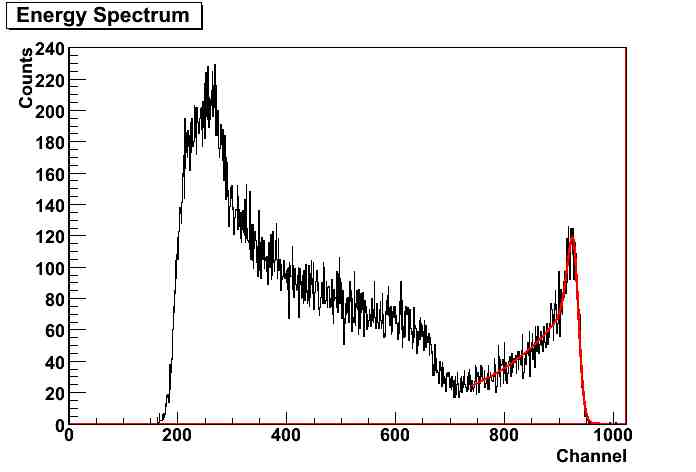}
\caption{Calibration spectrum taken with a Cs-137 source at 1.0kV bias with
  the Frisch collar detector \# 1 using a 9mm collar. Subsequently the
  resolution deteriorated (see table (1))
  probably due to
  handling the crystal over a long period in air for testing various collar
  design. Early on this detector shows a 1.33\% energy resolution. The fitted
  function is indicated as solid red line on top of the
  histogram.\label{middle1kV}}
\end{center}
\end{figure}
\par
\begin{table}[htb]
\begin{center}
\vspace*{6pt}
\begin{tabular}{lll}
Detector and conditions & Peak energy (keV) & FWHM (\%) \\\hline
CAPture; sum signal & 661.6 & 0.82 \\
CAPture; anode-only & 661.6 & 1.04 \\
CAPture; cathode irradiation & 356 & 1.09 \\
CAPture; anode irradiation & 356 & 1.14 \\
Frisch collar \#{}1, 800V, 9.5mm collar & 661.6 & 1.9 \\
\#{}1, 800V, 9.5mm, cathode irradiation & 356 & 2.7 \\
\#{}1, 800V, 9.5mm, anode irradiation & 356 & 3.0 \\
\#{}1, 1.5kV, 9mm collar & 661.6 & 1.81 \\
\#{}1, 1.0kV, 9mm collar & 661.6 & 1.33 \\
\#{}2, 800V, 9mm collar & 661.6 & 2.2 \\
\#{}3, 1kV, 9mm collar & 661.6 & 2.4
\end{tabular}
\caption{Energy resolution results obtained in this work. The outstanding
  results obtained with the CAPture detector are listed at the
  top. Subsequently, 
  the Frisch collar detectors are numbered as \#{}1,2,3. The bias and collar
  height show conditions under which the best results were
  obtained. Re-dressing detector \#{}1 with
  various new collars over time has worsened the resolution since its best
  result, the 1.33\% (see figure (10)), was obtained as one
  of the first measurements. The second result at 1.5kV is listed to indicate
  how over-biasing affects the result since the detector was not stable at this
bias.\label{tab1}}
\end{center}
\end{table}
\section{Conclusion}
A new readout and data analysis method for single channel, room-temperature
semiconductor detectors is introduced which significantly
improves energy resolution despite its simplicity. The new preamplifer circuit
and data analysis methods have been discussed in detail. The application
to three potentially rather poor CZT crystals operated as Frisch collar
detectors yields energy resolutions
comparable to or better than existing values from literature either for Frisch
collar detectors or coplanar grid detectors of similar size (but often, where
mentioned, far superior crystal quality). Our chosen test standard for
comparison, an eV-Products CAPture Plus detector, surpasses all expectations
and shows energy resolutions, to the best of our knowledge, better than any
measured with a 
similar device so far. Energy resolutions for
such a large volume CZT detector of well under 1\% at 662 keV without
subsequent 
corrections, i.e. efficiency losses, have so far only been
reported for much more complicated (in terms of fabrication and readout)
pixelated crystals.


\begin{thebibliography}{00}
\bibitem{cobra}K. Zuber, {\em Phys. Lett.} {\bf B519} (2001) 1; T. Bloxham et
  al., COBRA collaboration, {\em Phys. Rev.} {\bf C76} (2007) 025501
\bibitem{collar}D.S. McGregor et al., {\em Appl. Phys. Lett.} {\bf 72} (1998)
  792; 
  W.J. McNeil et al., {\em Appl. Phys. Lett.} {\bf 84} (2004) 1988;
  G. Montemont et al., {\em IEEE Trans. Nucl. Sci.} {\bf 48} (2001) 278;
  A. Kargar et al., 
  {\em Nucl. Instr. Meth.} {\bf A558} (2006) 497; A. Kargar et al.,
  {\em Nucl. Instr. Meth.} {\bf A562} (2006) 262; M. Harrison, A. Kargar and
  D.S. McGregor, {\em Nucl. Instr. Meth.} {\bf A579} (2007) 134
\bibitem{capture}D.S. Bale and C. Szeles, {\em Proc. SPIE} {\bf Vol. 6319}
  (2006); 
  C. Szeles et al., ibid. (2006) 191; see also White Papers at
  {\em http://www.evproducts.com/czt\_white\_papers.html} 
% \href{http://www.evproducts.com/czt\_white\_papers.html}{www.evproducts.com}
\bibitem{mixed}L. Verger et al., {\em IEEE Trans. Nucl. Sci.} {\bf 52} (2005)
  1733 
\bibitem{He}Zhong He et al., {\em Nucl. Instr. Meth.} {\bf A380} (1996) 228;
  {\em Nucl. Instr. Meth.} {\bf A388} (1997) 180
\bibitem{barrett}H. Barrett, J. Eskin and H. Barber, {\em Phys. Rev. Lett.}
  {\bf 75} (1995) 156
\bibitem{hamrita}A. Hamrita et al., {\em Nucl. Instr. Meth.} {\bf A531} (2004)
  607 
\bibitem{dsp}St. W. Smith, {\em The Scientist and Engineer's Guide to Digital
Signal Processing}, Cal. Tech. Publ. (1997) USA; see also
{\em http://www.DSPguide.com}
%\href{http://www.DSPguide.com}{www.DSPguide.com}
\bibitem{arlt}Gunnink and Arlt, {\em Nucl. Instr. Meth.} {\bf A458} (2001) 196
\bibitem{ieee}R.M. Keyser, {\em IEEE Nuclear Science Symposium Conf. Record}
  (2001) 315 
\bibitem{verger}L. Verger et al., {\em Nucl. Instr. Meth.} {\bf A571} (2007) 33
\end{thebibliography}
\end{document}